\documentclass[12pt]{iopart}

\usepackage{dcolumn}
\usepackage{bm}
\usepackage{graphicx}

\def\be{\begin{equation}}
\def\ee{\end{equation}}
\def\bea{\begin{eqnarray}}
\def\eea{\end{eqnarray}}
\def\bse{\begin{subequations}}
\def\ese{\end{subequations}}

\def\be{\begin{eqnarray}}
\def\ee{\end{eqnarray}}

\begin{document}

\title[]{Enhanced Raman sideband cooling of caesium atoms in a vapour-loaded magneto-optical trap}

\author{Y Li$^1$, J Wu$^{1,2}$, G Feng$^1$, J Nute$^2$, S Piano$^2$, L Hackerm\"{u}ller$^2$, J Ma$^1$, L Xiao$^1$ and S Jia$^1$}
\address{
$^1$State Key Laboratory of Quantum Optics and Quantum Optics Devices, Institute of Laser Spectroscopy,
College of Physics and Electronics, Shanxi University, Taiyuan 030006, China\\
$^2$School of Physics and Astronomy, University of Nottingham, UK\\}

\ead{wujz@sxu.edu.com and mj@sxu.edu.cn}
\vspace{10pt}
\begin{indented}
\item[]\today
\end{indented}

\begin{abstract}
We report enhanced three-dimensional degenerated Raman sideband cooling (3D DRSC) of caesium (Cs) atoms in a standard single-cell vapour-loading magneto-optical trap. Our improved scheme involves using a separate repumping laser and optimized lattice detuning. We load 1.5 $\times$ 10$^{7}$ atoms into the Raman lattice with a detuning of -15.5 GHz (to the ground F = 3 state). Enhanced 3D DRSC is used to cool them from 60 $\mu$K to 1.7 $\mu$K within 12 ms and the number of obtained atoms is about 1.2 $\times$ 10$^{7}$. A theoretical model is proposed to simulate the measured number of trapped atoms. The result shows good agreement with the experimental data. The technique paves the way for loading a large number of ultracold Cs atoms into a crossed dipole trap and efficient evaporative cooling in a single-cell system.
\end{abstract}

%

\section{Introduction}

Since the first experimental achievement of Bose-Einstein condensation (BEC) in dilute alkali atomic gases in 1995 [1-3], ultracold atoms have allowed the observation of Feshbach collision resonances [4, 5], the preparation of ultracold molecules in deeply bound states [6, 7] and several applications to quantum information processing [8, 9]. Evaporative cooling has become a well-established technique for achieving Bose-Einstein condensation of most alkali bosonic atoms [10]. The elastic collisions during the evaporative cooling rethermalize different kinds of alkali atoms that have been confined in magnetic or optical traps [1-3, 11]. The more collisions there are, the more quickly and efficiently the cooling process will occur. However, caesium's magnetically trappable states suffer from large inelastic scattering losses, and so, for a long time, it was thought impossible to condense this species [12-14]. Because of the special collisional characteristics of caesium atoms, an alternative approach towards obtaining the BEC is to use a high-field seeking state (F = 3, mF = 3) in which all inelastic two-body processes are endothermic and therefore are suppressed at sufficiently low temperature [15-17].
\par Degenerated Raman sideband cooling (DRSC) is an indispensable cooling stage that reaches the temperature of $\sim$1 $\mu$K and simultaneously spin-polarizes the atoms in the high-field seeking state, preparing the atoms for evaporative cooling in an optical trap [15-17]. DRSC was first proposed in one dimension by Chu in 1998 [18], and was extended to three dimensions specifically for the condensation of caesium [19-21]. This technique requires a large number of atoms to be loaded into the 3D Raman lattice and the lattice light frequency should be resonant with the F = 4 $\rightarrow$ F' = 4 transition such that it simultaneously acts as a weak repumping light to avoid population of the ground F = 4 state [15, 17, 20]. Results reported previously indicate that, in order to reach a large loading rate in the Raman lattice, the 3D DRSC must begin at an initial temperature as low as $\sim$10 $\mu$K in a Magneto-Optical Trap (MOT) [15-17].
\par In this paper, we demonstrate an enhanced 3D DRSC scheme for trapped caesium atoms in a standard single-cell vapour-loaded MOT with a relaxed starting temperature requirement of approximately 60 $\mu$K. Instead of using the lattice light as a repumper, we introduce a separate repumping laser which enables freedom of choice in terms of lattice detuning. We then optimize the frequency detuning of the Raman lattice light, while considering simultaneously both the loading potential of the 3D Raman lattice and the heating losses owed to the interaction between the lattice light and trapped atoms. We propose a theoretical model to characterize the experimental results. The dependence of the number of atoms loaded in the lattice on the frequency detuning presents good agreement with the model.

\section{Experimental setup}

The experimental setup is shown in figure 1. The cold atoms are confined in a standard single-cell vapour-loaded MOT with a background pressure of $\sim$2.5 $\times$ 10$^{-8}$ Pa [22]. The sample temperature is measured at $\sim$200 $\mu$K using time-of-flight (TOF) methods and the atom number is measured at 9 $\times$ 10$^{7}$ using absorption imaging. The MOT is compressed by increasing the magnetic field gradient to 30 G/cm over 25 ms and holding for 40 ms. Simultaneously the MOT beams are red detuned 45 MHz from the F = 4 to F' = 5 transition. Optical molasses cooling is performed for 2 ms during which time the magnetic field gradient is switched off and the MOT beams are further detuned to -70 MHz. After optical molasses, the atom number is $\sim$3 $\times$ 10$^{7}$ with a temperature of 60 $\mu$K, which is relatively high compared with the $\sim$10 $\mu$K reported by other groups [15-19]. The atoms are subsequently loaded into a 3D Raman lattice formed by four far-off-resonance laser beams; two counter-propagating beams along the y axis (beams 2 and 3 in figure 1), and two beams along each of the x and z axes (1 and 4 in figure 1). Using only four linearly polarized laser beams ensures stabilization of the Raman lattice [23]. The linear polarizations of the x and z beams lie in the x-z plane, whereas the y beams are polarized with $\pm$10 degrees to the line bisecting the x and z beams, respectively.
\begin{figure}[!htbp]
\begin{center}
\includegraphics[width=8cm]{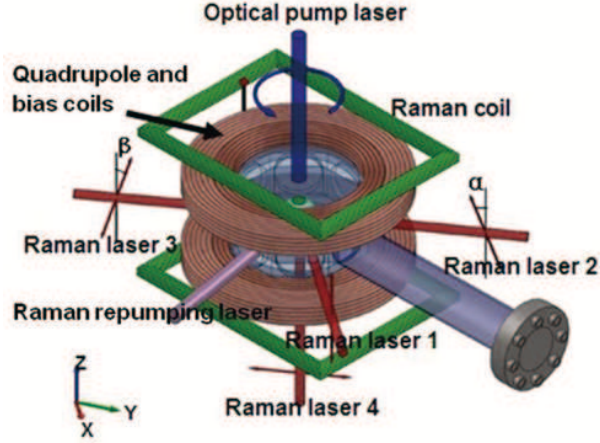}
\caption{\label{fig:epsart} Experimental setup for 3D Raman lattice and magnetic field configuration for 3D DRSC.}
\end{center}
\end{figure}
\par The Raman lattice light is switched on 0.5 ms before the end of the optical molasses stage. As the optical molasses stage ends, the optical pumping and repumping lasers are switched on to begin the cooling cycle which is held for 12 ms. We use a laser power of 25 mW for each Raman lattice laser beam with a radius of 1.15 mm (1/e$^{2}$) and an optimized red detuning of -15.5 GHz from the 6S$_{1/2}$, F = 3 $\rightarrow$ 6P$_{3/2}$, F = 4 transition. A pair of Helmholtz coils provide a magnetic field of $\sim$ 200 mG which brings two vibrational energy levels from different Zeeman hyperfine states into degeneracy. The mainly $\sigma$$^{+}$ polarized optical pumping laser is 8 MHz blue-detuned from the F = 3 $\rightarrow$ F' = 2 transition. It is derived from the same laser that provides the repumping laser for the MOT and has an intensity of 1 mW/cm$^{2}$. The optical pumping laser is tilted by just five degrees to generate a small amount of $¦Ð$ light along the magnetic field B in the x-y plane.

\section{Raman lattice}

\par A 3D Raman lattice is usually used after the optical molasses stage to reach an effective 3D DRSC of atoms. The lattice consists of four beams with linear polarization in the y-z plane. The four beams are of equal amplitude and the same frequency, which is red detuned to the F = 3 $\rightarrow$ F' = 4 transitions. The field of the lattice is written as
\begin{equation}
\mathbf{E}(\mathbf{r},t)=\emph{E}_{0} \mathbf{\bm{\varepsilon}}(\mathbf{r})e^{-i\omega_{L}t}+c.c.,
\end{equation}
\begin{equation}
\bm{\varepsilon}(\mathbf{r})=\mathbf{n}_{1}e^{ikx}+\mathbf{n}_{2}e^{iky}+\mathbf{n}_{3}e^{-iky}+\mathbf{n}_{4}e^{ikz},
\end{equation}
where \emph{E}$_{0}$ is the amplitude of light field and $\mathbf{n}$$_{i}$ is the unit polarization vector of the $\emph{i}$th lattice beam light [24] expressed as
\begin{equation}
\mathbf{n}_{1}=\left(\begin{array}{c}0\\ 0\\ 1\end{array}\right),
\mathbf{n}_{2}=\left(\begin{array}{c}\cos\alpha\\ 0\\ \sin\alpha\end{array}\right),
\mathbf{n}_{3}=\left(\begin{array}{c}\cos\beta\\ 0\\ \sin\beta\end{array}\right),
\mathbf{n}_{4}=\left(\begin{array}{c}1\\ 0\\ 0\end{array}\right),
\end{equation}
where $\alpha$ and $\beta$ are the intersection angles of the polarization direction for lattice beams 2 and 3 with z axis, respectively. The optical potential for atoms in the ground F = 3 state is given as
\begin{equation}
\widehat{\mathbf{U}}(\mathbf{r})=-\frac{2}{3}U_{0}| \mathbf{\bm{\varepsilon}}(\mathbf{r})|^{2}\widehat{\mathbf{I}}
    +\frac{i}{3}U_{0}[\mathbf{\bm{\varepsilon}}(\mathbf{r})^{\ast} \times \mathbf{\bm{\varepsilon}}(\mathbf{r})]\cdot\frac{\widehat{\mathbf{F}}}{F},
\end{equation}
where $\widehat{\mathbf{I}}$ and $\widehat{\mathbf{F}}$ are the identity and angular momentum operators. The first term in Eq.(4) is the energy shift that is proportional to the energy density of optical field and identical for all Zeeman sub-levels. The second term indicates an effective magnetic field, the direction of which is determined by vector $i[\mathbf{\bm{\varepsilon}}(\mathbf{r})^{\ast} \times \mathbf{\bm{\varepsilon}}(\mathbf{r})]$ [24, 25]. The magnitude of the light shift induced by a single lattice beam is expressed as U$_{0}$=$\frac{\hbar\Gamma^{2}}{\Delta}$$\frac{I}{8I_{s}}$, where $\Gamma$ = 2$\pi$ $\times$ 5.28 MHz is the natural line width, $\Delta$ is the lattice beam frequency detuning from the ground 6S$_{1/2}$, F=3 state, $I$ is the intensity of beam light, and $I$$_{s}$ = 1.1 mW/cm$^{2}$ is the saturation intensity for caesium atoms.

\begin{figure}[!htbp]
\begin{center}
\includegraphics[width=8cm]{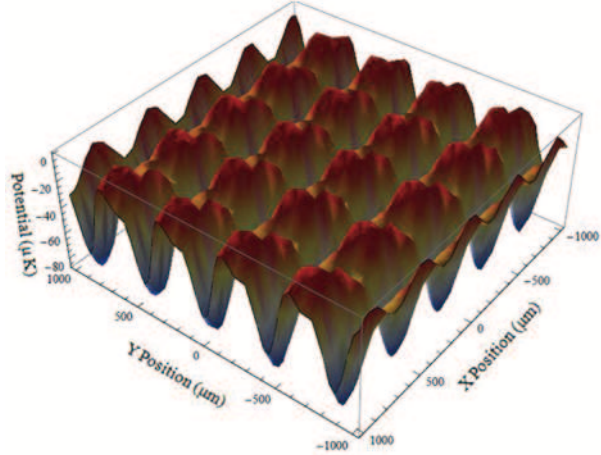}
\caption{\label{fig:epsart} Potential of 3D Raman lattice that consists of four linear polarization lasers, two counter-propagating beams along the y axis with the polarization of $\pm$10 degrees to the line bisecting the x and z beams and the other two beams along each of the x and z axes with the linear polarizations in the x-z plane as shown in figure 1.}
\end{center}
\end{figure}

\par The vibrational frequency of the 3D Raman lattice is crucial for determining the magnetic field used in 3D DRSC, so that the Zeeman splitting between two adjacent sub-levels with $\Delta$mF = 1 well matches the spacing of the vibrational levels in the lattice. The lattice potential according to Eq.(4) is shown in figure 2, where the detailed parameters can be found in the experimental part and the scheme describing directions and polarizations of the laser beam have been shown figure 1. The corresponding vibrational frequencies are $v$$_{x}$ = 86.1 kHz, $v$$_{y}$ = 48.2 kHz and $v$$_{z}$ = 48.2 kHz. After CMOT and optical molasses, the atoms are prepared on the ground F=3 state and the corresponding Land$\acute{e}$ factor g is $-$ 0.2514. Taking into account the splitting of the Zeeman sub-levels in the F = 3 mainfold, a theoretical value of 180 mG is in good agreement with the optimal magnetic field used in our experiment.

\section{Optimization for frequency detuning of Raman lattice}

In order to suppress losses owed to atoms being off-resonantly pumped by the lattice light and decaying to the ground F = 4 state, a weak repumping laser ($\sim$30 $\mu$W) acting on the 6S$_{1/2}$, F = 4 $\rightarrow$ 6P$_{3/2}$, F' = 3 transition is used. This light is derived from the same laser diode that provides trapping lasers in the MOT and so we can independently optimize the frequency detuning. As a result, we can effectively trap and hold more atoms in the ground F = 3 state starting at a relatively high temperature of optical molasses.
\par The dependence of the atom number in the lattice on detuning is shown in figure 3(a). The number of atoms in the lattice initially increases with detuning because larger detuning reduces heating losses from the interaction between the lattice light and the atomic sample. As the red detuning increases further, however, the potential depth drops and so the atomic loading rate also drops. In order to understand the physical mechanism and find the optimal value of the detuning precisely, a theoretical analysis is briefly introduced here. In this model the atom number is determined by the loading potential of the Raman lattice and the interaction with the trapped atoms. Thus the detuning frequency has a great impact on the number of trapped atoms. We define the number of atoms trapped in the lattice by introducing two parameters, $\gamma$ and $\eta$, as
\begin{equation}
N=\gamma \times 1.965 \times U_{0} + \eta \times \sigma(\omega),
\end{equation}
where $1.965 \times U_{0}$ is the loading potential of the Raman lattice and $\sigma(\omega)$ is the cross-section that characterizes the atomic loss during the interaction with the lattice beam. For a two-level atom, the cross-section of the atoms in the 6S$_{1/2}$, F = 3 state is given as
\begin{equation}
\sigma(\omega) = \frac{\sigma_{0} \Gamma^{2}/4}{(\omega - \omega_{0})^{2} + \Gamma^{2}/4},
\end{equation}
where $\omega$ is the frequency of the lattice light, $\omega_{0}$ is the transition frequency from the 6S$_{1/2}$, F = 3 state to the 6P$_{3/2}$, F = 4 state and $\sigma_{0}$ = $\sigma(\omega_{0})$ is defined as the maximum cross-section [26]. The number of atoms trapped in the 3D Raman lattice can be expressed as a function of lattice detuning $\Delta$:
\begin{equation}
N=\gamma \times \frac{1.965 \hbar \Gamma^{2} I}{8 I_{s} \Delta} + \eta \times \frac{\sigma_{0}}{1 + 4 \Delta^{2}/\Gamma^{2}}.
\end{equation}
This formula has been applied to fit the experimental result as shown in figure 3(a). It is evident that it is reasonable to use a red detuning of -15.5 GHz for the Raman lattice light in our experiment. We obtain 1.5 $\times$ 10$^{7}$ atoms in the ground F = 3 state in the lattice. The spatial distribution and integrated optical density of the cold atomic cloud are illustrated in figure 3(b) and (c).
\begin{figure}[!htbp]
\begin{center}
\includegraphics[width=8cm]{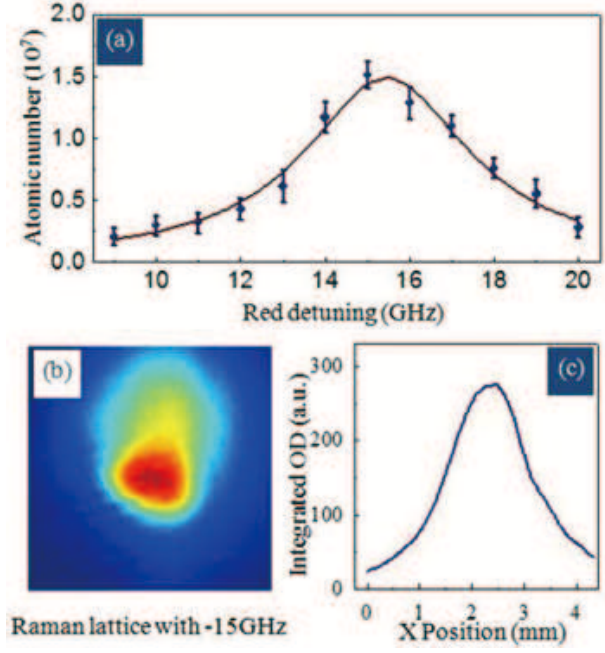}
\caption{\label{fig:epsart} (a) Variation of the number of atoms loaded into the Raman lattice after optical molasses as a function of lattice laser detuning from the 6S$_{1/2}$, F = 3 state. The highest loading rate of $\sim$1.5 $\times$ 10$^{7}$ atoms is obtained at -15.5 GHz. The solid curve represents a fitting line according to Eq. (7). (b) Absorption image of atoms loaded into the lattice taken after a flight time of 5 ms. (c) The integrated optical density corresponding to figure (b) along the vertical direction.}
\end{center}
\end{figure}

\section{Experimental results}

The 3D Raman lattice successfully isolates the atoms into individual lattice sites [19-21]. We optimize the magnetic field and find that 200 mG is beneficial to degenerate $|$ F = 3, mF = 3; $v$ $\rangle$, $\mid$ F = 3, mF - 1; $v$ - 1 $\rangle$ and $\mid$ F = 3, mF - 2; $v$ - 2 $\rangle$ states by Raman coupling from lattice beams. $v$ and mF denote the vibrational quantum number and hyperfine sub-level respectively. The optical pumping takes place in the 6S$_{1/2}$, F = 3 $\rightarrow$ 6P$_{3/2}$, F' = 2 transition. $v$ is conserved in the Lamb-Dicke regime and thus leads to a rapid loss of the vibrational quantum number. The atoms then populate at the vibrational ground $\mid$ F = 3, mF = 2; $v$ = 0 $\rangle$ state, which is not resonant with the $\sigma$$^{+}$ polarized beam and simultaneously depopulated by a weak $\pi$ polarized component from the beam. In the end, all atoms are transferred to the dark $\mid$ F = 3, mF = 3; 0 $\rangle$ state.

\par After 12 ms of cooling, the optical pumping and repumping lasers are switched off while the lattice light is turned down adiabatically according to p(t) = P(0) $[$ 1 + t/t$_{0}$ $]$$^{-2}$ in 500 $\mu$s, where t$_{0}$ is typically defined at 100 $\mu$s [19, 20]. Then, 1.2 $\times$ 10$^{7}$ atoms spin-polarized in the F = 3, mF = 3 state at a density of 10$^{10}$ cm$^{-3}$ remain after cooling. The spatial distribution of atoms is shown in figure 4(a) where the absorption image is taken horizontally after 10 ms TOF. The Gaussian fitting has been performed both in the vertical and in the horizontal direction as shown in figure 4(b) and figure 4(c), respectively. The temperature of the ensemble is measured at 1.7 $\mu$K using the TOF method shown in figure 4(d).

\begin{figure}[!htbp]
\begin{center}
\includegraphics[width=8cm]{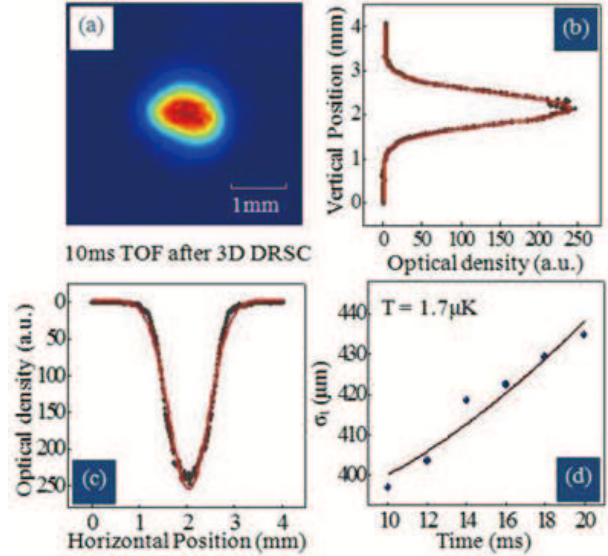}
\caption{\label{fig:epsart}  (a) Absorption image taken from the horizontal direction after 10 ms TOF. (b) and (c) are Gaussian fitting results along the vertical direction and horizontal direction respectively. The dots are experimental data, and the red solid lines are the Gaussian fitting results. (d) The dots are the expansion of atomic cloud after the release of 3D Raman lattice; the solid line is the fitting result corresponding to a temperature of 1.7 $\mu$K.}
\end{center}
\end{figure}

\par During 3D DRSC atoms that fall to the F = 4 ground state need be pumped back to the ground F = 3 state, so often the lattice frequency is set at 9.2 GHz red detuned from the F = 3 $\rightarrow$ F' = 4 transition so that the lattice light also acts as a repumper [15, 17, 20] . However, as shown in figure 3(a), this frequency detuning of the lattice beam is not optimal for the high initial temperature of $\sim$60 $\mu$K because of the large and rapid loss of the atoms in the Raman lattice. One can clearly see that a larger red detuning of -15.5 GHz greatly facilitates loading a large number of atoms into the lattice. The optimization of the frequency detuning of the lattice beam allows us to obtain an enhanced DRSC in a 3D Raman lattice for caesium atoms.

\section{Conclusion}
In conclusion, we have performed 3D DRSC on trapped caesium atoms in a vapour cell starting from a high initial temperature of $\sim$60 $\mu$K. We begin with 3 $\times$ 10$^{7}$ atoms and finish with 1.2 $\times$ 10$^{7}$ atoms spin-polarized in the F = 3, mF = 3 state. At the end of the cooling stage, we obtain a temperature of 1.7 $\mu$K. Taking into account the loading potential of the 3D Raman lattice and its heating loss for the trapped atoms, we propose a theoretical model to simulate the atom numbers loaded into the lattice. The numbers at different frequency detuning are given, resulting in a good agreement with the theoretical analysis. As a result, the detuning of the Raman lattice light is optimized to avoid large and rapid loss of the caesium atoms at a relatively high temperature. Our scheme's greatest advantage is the relaxed starting temperature requirement, meaning that 3D DRSC can be directly accomplished in a standard single cell vapour-loaded MOT without reliance on any additional devices such as a Zeeman slower. Furthermore, only conventional red detuned optical molasses has been performed in the ground F = 4 state, which is relatively simple compared with the blue Sisyphus cooling in the ground F = 3 state [18]. This simple and robust scheme can be adapted to a number of other atomic species. The enhanced 3D DRSC paves the way for effective loading of the cold atomic sample into the optical dipole trap and following evaporation cooling to obtain the Cs BEC.

\section*{Acknowledgments}
This work was supported by 973 Programme (No. 2012CB921603), PCSIRT (No. IRT13076),  NSF of China (No. 91436108, No. 61378014, No. 61308023, No. 61275211, No. 61378015 and No. 11434007), SRFDPHE (No. 20131401120012), NFFPT (No. J1103210) and NSFSXP (No. 2013021005-1). J.W. acknowledges the support from the Royal Society K. C. Wong Postdoctoral Fellowship.

\end{document}